\documentclass[aps,prl,twocolumn,superscriptaddress]{revtex4-1}
\usepackage{graphicx}  
\usepackage{natbib}
\begin{document}

\title{Phase diagram and effective shape of semi-flexible colloidal
  rods and biopolymers}

\author{M.~Dennison} 
\author{M.~Dijkstra}
\affiliation{Soft Condensed Matter, Debye Institute for Nanomaterials
  Science, Utrecht University, Princetonplein 5, 3584 CC Utrecht, The
  Netherlands}
\author{R.~van Roij} 
\affiliation{Institute for Theoretical Physics, Utrecht University,
  Leuvenlaan 4, 3584 CE Utrecht, The Netherlands}

\date{\today}

\begin{abstract}
We study suspensions of semi-flexible colloidal rods and biopolymers
using an Onsager-type second-virial functional for a segmented-chain
model. For suspensions of thin and thick fd virus particles we
calculate phase diagrams in quantitative agreement with experimental
observations, and we find their effective state-point dependent shape
to be much shorter and thicker than the actual shape. We also
calculate the stretching of worm-like micelles in a host fd virus
solution, again finding agreement with experiments. For both systems,
our results show that the fd virus stiffness can play a key role in
system behavior.
\end{abstract}

\maketitle

Rod-like particles are capable of forming a great variety of
liquid-crystalline phases \cite{ref:Chandrasekhar}, and have been
widely explored experimentally and theoretically
\cite{ref:Experi1}. Recently a prominent role is being played by
aqueous suspensions of fd virus particles, which are charged,
semi-flexible, Brownian needles with a length-to-diameter ratio
exceeding 100, exhibiting isotropic (I), cholesteric nematic (N),
smectic, columnar, and crystalline phases upon increasing the
concentration
\cite{ref:Tang_Fraden,ref:Dogic_Fraden,ref:grelet_PRL100_168301_2008}. Moreover,
wild-type fd virus particles have been bio-engineered to have, for
instance, a polyethylene-glycol (PEG) coating, such that mixtures of
thin and thick rods with diameter ratios $d$ varying from $1.1$ to
$3.7$ could be studied experimentally \cite{ref:Purdy}. The resulting
experimental phase diagrams are extremely rich, even in the regime
where only I and N phases are relevant. For instance, the observations
not only include I-N coexistence with strong fractionation effects,
but also, for $d\gtrsim 3$, two-phase N-N and three-phase I-N-N
coexistence with phase diagram topologies that strongly depend on
$d$. Given the needle-like shape of the fd virus particles, and the
relative structural simplicity of I and N phases, one would expect
these phase diagrams to be well understood, e.g. resembling those of
theoretical predictions based on Onsager's second virial theory for
thin-thick mixtures of rigid rods \cite{ref:vRoij}. However, these
systems turn out to be surprisingly poorly understood. For instance,
the experiments show N-N demixing with a low-density (lower) critical
point that shifts to lower densities with increasing $d$
\cite{ref:Purdy}, while the theory predicts the exact opposite: a
high-density (upper) N-N critical point and N-N demixing that extends
to higher densities with increasing $d$ \cite{ref:vRoij}. Moreover,
the experiments show N-N demixing at diameter ratios as low as
$d\gtrsim 2-3$ while rigid-rod models predict $d\gtrsim 4-5$
\cite{ref:vRoij,ref:Purdy}. In this Letter we will show that the key
to a real understanding of these systems is {\em flexibility}, which
renders the needles effectively shorter and fatter depending on the
state point
\cite{ref:Fynewever,ref:semenov_and_subbotin_polymerscience31_2266_1989}. In
addition, our model and theory can also quantify the observed {\em
  stretching} of guest biopolymers in host suspensions of fd virus
particles \cite{ref:Dogic_host}. 

One-component systems of semi-flexible rods have been studied using
numerous methods, and it is known that only a slight flexibility is
enough to shift the I-N transition to significantly higher
concentrations
\cite{ref:KS1,*ref:KS2,ref:Chen,ref:Dijkstra_Frenkel,ref:Odijk,ref:wessels1,*ref:wessels2}. We
build on the segmented-chain model introduced by Wessels and Mulder
\cite{ref:wessels1,*ref:wessels2}, in which (i) flexibility is
incorporated by introducing a bending potential between the chain
segments, and (ii) excluded volume is taken into account at the
segment level. This approach reproduces (in the appropriate limits)
the results of Ref.~\cite{ref:KS1,*ref:KS2}, with the advantage of
only having to deal with a discrete number of degrees of freedom,
making explicit calculations of extensions to two-component systems
feasible.

We consider a suspension of $N_i$ semi-flexible rods of species
$i=1,2$ with contour lengths $L_i$, in a volume $V$ at temperature
$T$. Following Ref.~\cite{ref:wessels1,*ref:wessels2} we model a rod
of species $i$ as a chain of $M_i$ rod-like segments of length
$l_i=L_i/M_i$ and diameter $D_i\ll l_i$. Denoting the orientation of
the $m$-th segment by a unit-vector ${\bf\omega}_m$ (with $1\leq m\leq
M_i$), we write the bending energy of a chain of species $i$ with
orientation ${\bf\Omega}=\{{\bf\omega}_1,\dots,{\bf\omega}_{M_i}\}$ as
\begin{equation}
U_{i}({\bf\Omega})=\sum_{m=1}^{M_{i}-1}u_{i}({\bf\omega}_{m},{\bf\omega}_{m+1})=
-\frac{P_i}{l_i}\sum_{m=1}^{M_{i}-1}{\bf\omega}_{m}\cdot{\bf\omega}_{m+1},
\label{eq:U}
\end{equation}
where the stiffness is described in terms of the persistence length
$P_i$ \cite{ref:wessels1,*ref:wessels2}. Here and below we use thermal
energy units by setting $k_BT=1$. The state of the suspension is
characterized by the orientation distribution functions (ODFs)
$f_i({\bf\Omega})$, which satisfy the normalization condition $\int
d{\bf\Omega} f_i({\bf\Omega})=1$ where $d{\bf\Omega}=
\prod_{m=1}^{M_i}d{\bf\omega}_m$. Denoting the total number of rods by
$N=N_1+N_2$, the density by $\rho=N/V$, and the mole fraction of
species $i$ by $x_i=N_i/N$, we can write the variational free-energy
functional $F[f_1,f_2]$ within an Onsager-like second virial
approximation as

\begin{eqnarray}
\frac{F}{N} &= &\ln(B\rho)-1+ x_{1}\ln x_{1}+ x_{2}\ln
x_{2} \label{eq:helmholtz_bi1}
\\ &+&\displaystyle\sum_{i=1}^{2}x_{i}\displaystyle \int
f_{i}({\bf\Omega})\Big(\ln(4\pi
f_{i}({\bf\Omega}))+U_{i}({\bf\Omega})\Big)d{\bf\Omega} \nonumber
\\ &+&\frac{\rho}{2}\displaystyle\sum_{i,j=1}^{2}x_{i}x_{j}\displaystyle
\int f_{i}({\bf\Omega})f_{j}({\bf\Omega}^{\prime}) K_{ij}({\bf
  \Omega},{\bf\Omega}^{\prime})d{\bf\Omega}d{\bf
  \Omega}^{\prime}.\nonumber
\end{eqnarray}
The first line of Eq.~(\ref{eq:helmholtz_bi1}) represents the
translational and the mixing ideal-gas contributions (with
$B=\frac{\pi}{4}D_{1}L_{1}^{2}$, a constant), the second line, the
orientation entropy and bending energy, and the third line the
excluded volume interactions, which can be considered at the segment
level with $K_{ij}({\bf
  \Omega},{\bf\Omega}^{\prime})=\sum_{m=1}^{M_{i}}\sum_{m^{\prime}=1}^{M_{j}}k_{ij}({\bf
  \omega}_{m},{\bf\omega}_{m^{\prime}})$. The free-energy functional
in Eq.~(\ref{eq:helmholtz_bi1}) is a two-component generalization of
the one-component segmented-chain functional of
Ref.~\cite{ref:wessels1,*ref:wessels2}, and for $M_i=1$ and $U_i\equiv
0$ it reduces to the Onsager functional for binary mixtures of rigid
rods \cite{ref:Vroege,ref:vRoij}. From hereon, we simply summarize our
method, and direct those interested in the complete outline to the
appendix. At a given thermodynamic state point, the equilibrium ODFs
minimize $F$ and therefore satisfy
\begin{eqnarray}
f_i({\bf\Omega})&=&\frac{\exp(-U_i({\bf\Omega})-V_i({\bf\Omega}))}{Q_i}; \label{sca}\\ 
V_i({\bf\Omega})&=&\rho\sum_{j=1}^2 x_j\int
K_{ij}({\bf\Omega},{\bf\Omega}')f_j({\bf\Omega}')d{\bf\Omega}',
\label{scb}
\end{eqnarray}
where $V_i({\bf\Omega})$ can be seen as a self-consistent field acting
on all segments of a chain, and $Q_i$ is a partition function-like
normalization factor \cite{ref:wessels1,*ref:wessels2}. Explicitly
solving Eqs.~(\ref{sca}) and (\ref{scb}) for state points of interest
would be prohibitively expensive computationally because of the
high-dimensional angular ${\bf\Omega}$-grids that would be required in
the case when $M_i\gg 1$. Instead, we formally evaluate the functional
$F$ of Eq.~(\ref{eq:helmholtz_bi1}) in its minimum by inserting the
solutions $f_i$ of Eqs.~(\ref{sca}) and (\ref{scb}) to find the
equilibrium free energy
\begin{eqnarray}
\label{eq1:helmholtz_bin_eq}
\displaystyle\frac{F_{\mbox{\small eq}}}{N}&&= \ln(B\rho) -1 +
x_{1}\ln\frac{x_{1}}{Q_{1}} + x_{2}\ln\frac{x_{2}}{Q_{2}} -
\frac{1}{2}\rho\sum_{i,j}^{2}x_{i}x_{j} \nonumber\\ 
&\times& \sum_{m=1}^{M_{i}}\sum_{m'=1}^{M_{j}} \int
k_{ij}({\bf\omega},{\bf\omega}')f_{i,m}({\bf\omega})f_{j,m'}({\bf\omega}^{\prime})d{\bf\omega}d{\bf\omega}^{\prime},
\end{eqnarray}
where $f_{i,m}({\bf\omega})$ is the ODF of the $m$-th segment
($m=1,\dots,M_i$) of a chain of species $i=1,2$ given by
$f_{i,m}({\bf\omega}) =\int
f_i({\bf\Omega})\delta({\bf\omega}-{\bf\omega}_{m})d{\bf\Omega}$. Eq.~(\ref{eq1:helmholtz_bin_eq})
implies that the thermodynamics does {\em not} require the full
solutions $f_i({\bf\Omega})$ but only the $M_i$ single-segment
distributions $f_{i,m}({\bf\omega})$ and the normalization factors
$Q_i$. The calculation of these is given in the appendix. With the
ODFs, and hence $F_{\mbox{\small eq}}$, known, all thermodynamic
properties such as osmotic pressure $\Pi$ and phase diagrams follow
\cite{ref:vRoij}.

We have calculated the phase diagrams for mixtures of bare fd
particles (species $1$, thin) and PEG-coated ones (species $2$,
thick), with equal contour and persistence lengths,
$L_{1}=L_{2}=0.88\mathrm{\mu m}$ and $P_{1}=P_{2}=2.2\mathrm{\mu m}$
\cite{ref:Purdy}. The bare fd diameter is fixed to
$D_{1}=6.6\mathrm{nm}$, and following the experiments of
Ref.~\cite{ref:Purdy} we consider several diameter ratios $d=D_2/D_1$
to describe varying thicknesses of the PEG coating. Throughout, we use
sufficient segments per virus particle such that we are in the
continuum limit for all state points of interest (see the
appendix). In Fig.~\ref{fig:PD_fd1} we show, for several $d$, the
resulting phase diagrams in the $(\eta_1,\eta_2)$ representation,
where $\eta_i$ denotes the packing fraction of species $i=1,2$. For
all $d$ we find isotropic-nematic (I-N) coexistence, with the
tie-lines indicating, for increasing $d$, an increasing fractionation
of the thinner and thicker rods preferentially into the I and N phase,
respectively.
\begin{figure}[h]
    \includegraphics[width=0.99\columnwidth,height=0.99\columnwidth,angle=270]{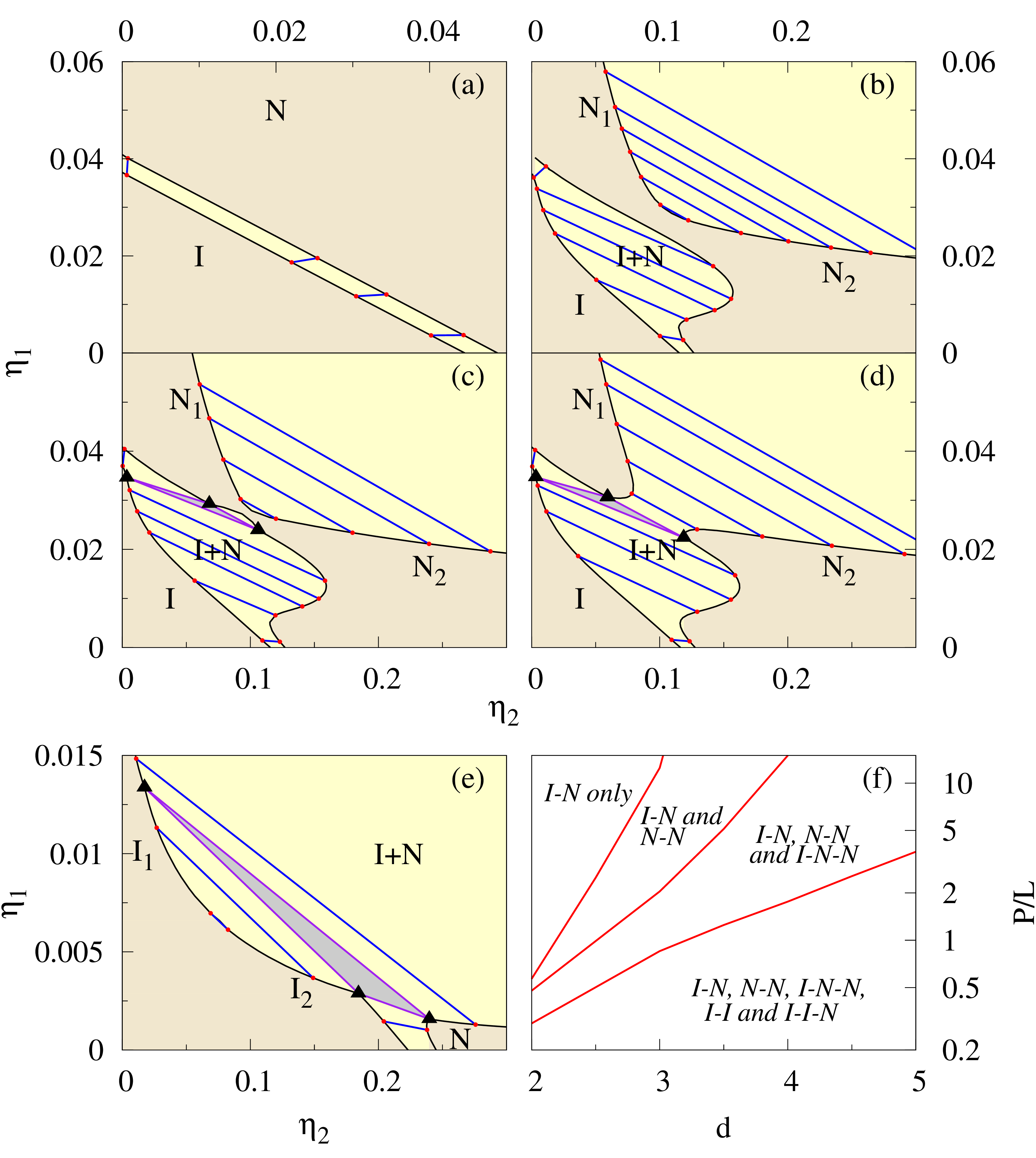}
    \caption{Phase diagrams (see text) for mixtures of bare (thin) fd
      virus particles (species 1) and PEG-coated (thick) ones (species
      2), with diameter ratios (a) $d=1.2$, (b) $3.1$, (c) $3.11$, (d)
      $3.125$, (e) $6$.  The lighter colored areas indicate the
      two-phase regions with tie-lines connecting coexisting
      state-points; triangles denote I-N-N and I-I-N triple
      points. (f) Topology of the phase diagram (two- and three phase
      coexistence) of binary mixtures of (modified) fd virus particles
      as a function of their diameter ratio $d$ and persistence length
      $P$.}
    \label{fig:PD_fd1}
\end{figure}

In agreement with the experiments of Ref.~\cite{ref:Purdy} we find
{\em no} N-N coexistence in the density regime of interest for the
smallest diameter ratio $d=1.2$ (Fig.~\ref{fig:PD_fd1}(a)). For
increasing $d$ an N-N demixing regime appears in the density regime of
interest, with a {\em lower} critical point as shown for $d=3.1$ in
Fig.~\ref{fig:PD_fd1}(b); in the experiments of Ref.~\cite{ref:Purdy}
such a phase diagram was found for $d\simeq 2.9$. Slightly increasing
the PEG layer thickness to $d=3.11$ reveals an I-N-N triple point and
an associated N-N demixing regime with an {\em upper} critical point
emerging out of the I-N coexistence regime
(Fig.~\ref{fig:PD_fd1}(c)). This N-N upper critical point has not been
reported experimentally, perhaps because it only exists in a small
regime of diameter ratios: for $d=3.125$ (Fig.~\ref{fig:PD_fd1}(d))
the upper and lower N-N points have merged to form a single
neck-shaped N-N regime that is separated from the I-N coexistence
regime by an I-N-N triple point, with the I and one N phase rich in
bare fd-particles and the other N phase rich in PEG-coated
fd-particles. Interestingly, such neck-shaped phase diagrams have {\em
  also} been reported in Ref.~\cite{ref:Purdy} for diameter ratios
$d\simeq3$. For $d=6$, Fig.~\ref{fig:PD_fd1}(e) shows an I-I-N triple
point and I-I coexistence, as found also for thin-thick mixtures of
rigid rods with diameter ratios exceeding $d\simeq8$
\cite{ref:vRoij}. We find that this behavior is present for
$d\gtrsim4.5$.

Motivated by recent progress in the bio-engineering of fd virus
particles \cite{ref:fd_changeP}, which may allow for tuning their
flexibility, we have also calculated binary-mixture phase diagrams for
a large variety of diameter ratios and persistence lengths
$P_1=P_2\equiv P$. Fig.~\ref{fig:PD_fd1}(f) summarizes our findings by
dividing the $(d,P)$ plane into regimes with phase diagrams featuring
only an I-N transition (for stiff rods and small $d$) all the way to
complex phase diagrams with I-N, I-I, N-N phase coexistence and I-I-N
and I-N-N triple points (for flexible rods and large $d$). Clearly,
increasing $d$ and decreasing $P$ have similar effects on the phase
diagram, and hence increasing the flexibility is expected to
considerably enhance the complexity of the phase diagrams.

In order for a model system of rigid rods to fully capture the phase
behavior of binary fd virus systems, one must use a system of much
shorter, thicker rods, with $L/D\lesssim7$ for the PEG-coated fd virus
\cite{ref:Purdy}, much lower than the true values of $20-110$. The
inference is that long semi-flexible rods exhibit the same phase
behavior as short rigid rods
\cite{ref:Fynewever,ref:semenov_and_subbotin_polymerscience31_2266_1989}. Our
model also enables us to study the effective shape exhibited by the
flexible rods throughout the binary fd virus systems. We define the
mean square effective length $L^{2}_{e,i}$ as
\begin{equation}
L^{2}_{e,i} = l^{2}_{i}\displaystyle
\sum_{m=1}^{M_{i}}\sum_{m^{\prime}=1}^{M_{i}}\int
({\bf\omega}\cdot{\bf\omega^{\prime}})f_{i,m,m^{\prime}}({\bf\omega},
{\bf\omega}^{\prime})d{\bf\omega}d{\bf\omega^{\prime}},
\label{eq:eff_length}
\end{equation}
where $f_{i,m,m^{\prime}}({\bf\omega},{\bf\omega}^{\prime})$ is the
pair orientation distribution function (PDF), the calculation for
which is given in the appendix. We use the PDFs to calculate $L_{e,i}$
and, from this, the effective diameter $D_{e,i}$ required for rigid
rods of length $L_{e,i}$ to have the same excluded volume as flexible
rods of length $L_{i}$.
\begin{figure}[h]  
  \includegraphics[width=0.4\columnwidth,height=0.99\columnwidth,angle=270]{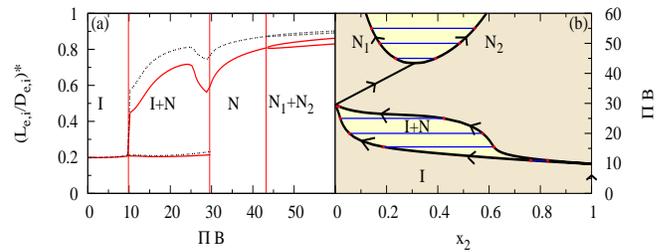}
  \caption{(a) Relative effective shape $(L_{e,i}/D_{e,i})^{*}$ $=$
    $(L_{e,i}/D_{e,i})/(L_{i}/D_{i})$ of fd virus particles as a
    function of osmotic pressure ($\Pi$) (solid lines show bare fd,
    dashed lines show PEG-coated fd), and (b) the path we follow
    throughout the phase diagram in the $x_{2}\;-\;\Pi$ representation,
    indicated by the arrows.}
  \label{fig:fd_length}
\end{figure}

Fig.~\ref{fig:fd_length}(a) shows the effective shape of the rods
$L_{e,i}/D_{e,i}$, for a mixture of thick-thin fd virus particles with
$d=3$, throughout the phase diagram (the route we follow is shown in
Fig.~\ref{fig:fd_length}(b), where we use the $x_{2}\;-\;\Pi$
representation for clarity). What is immediately apparent is that
throughout the phase diagram, while the rods always behave as shorter,
thicker rods, the effective shape varies considerably. In the
isotropic phase, we find $L_{e,i}/D_{e,i}$ to be about $20\%$ of
$L_{i}/D_{i}$ (for both species). This corresponds to
$L_{e}/D_{e}\approx8.5$ for the thick rods, close to the
$L/D\lesssim7$ required for rigid rod systems to capture the phase
behavior of these binary systems \cite{ref:Purdy}. For the nematic
phase, however, $L_{e,i}/D_{e,i}$ jumps to over $50\%$ of
$L_{i}/D_{i}$, and increases considerably as $\Pi$ is increased. We
conclude that a fixed effective shape does not capture the essential
physics of these suspensions; the state-point dependent stretching of
the flexible rods is a key feature.

Finally, we present results for the effective length of semi-flexible
polymers dissolved in an fd virus suspension. A range of polymers
which undergo a coil-rod transition, stretching out over the I-N
transition of the host fd virus, has been studied experimentally
\cite{ref:Dogic_host}. Here, we examine worm-like micelles, which have
constant $P=0.5\mathrm{\mu m}$, $D=14\mathrm{nm}$ and variable
$L=5-50\mathrm{\mu m}$. The concentration of the polymers is
sufficiently low that they can be treated as a single particle in a
bulk fd virus suspension, and we study the behavior of the polymer
$L_{e}$ over the I-N phase transition of the fd virus (which is at
$\Pi B=29.54$). The results are shown in Fig.~\ref{fig:f-actin_fd
  virus}(a). For the shortest polymers studied, we see a considerable
jump in $L_{e}$, from $L_{e}\simeq0.26L$ to $\simeq0.61L$,
corresponding to a coil-rod transition. For longer polymers, the jump
in $L_{e}$ becomes smaller, and the longest ones only become truly
rod-like well into the nematic phase of the fd virus. It is
interesting to note that in the isotropic phase, for all cases, $L_{e}$
appears to remain essentially constant.
\begin{figure}[h]
  \includegraphics[width=0.675\columnwidth,height=0.99\columnwidth,angle=270]{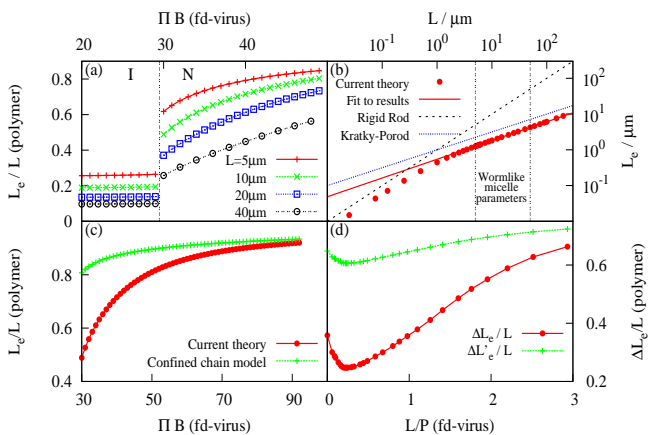}
  \caption{(a) $L_{e}$ of worm-like micelles of various $L$ against
    host fd virus osmotic pressure $\Pi$. (b) $L_{e,I}$ against $L$,
    with fit to $L_{KP}$, compared to the ideal Kratky-Porod $L_{KP}$
    and rigid rod $L$. (c) $L_{e,N}$ of worm-like micelles with
    $L=10\mathrm{\mu m}$, and $L_{O}$ of a confined semi-flexible
    polymer, against $\Pi$. (d) $\Delta L_{e}$ of worm-like micelles
    against $L/P$ of host fd virus, compared to $\Delta
    L^{\prime}_{e}$}
  \label{fig:f-actin_fd virus}
\end{figure}

In the isotropic fd virus phase, the polymer may be considered along
the lines of the Kratky-Porod worm-like chain model
\cite{ref:Kraty_Porod}, generalized to account for excluded volume
effects \cite{ref:Binder}. The average end-to-end length is defined as
$L_{KP}=\sqrt{4P^{\prime}P}(L/2P)^{\nu}$, where $P^{\prime}$ is an
effective persistence length. For an ideal Kratky-Porod chain
$P^{\prime}=P$ and $\nu = 0.5$. We calculate $L_{e}$ in the isotropic
phase ($L_{e,I}$) for a large range of $L$ values, and fit $L_{KP}$ to
our results for the range of worm-like micelle parameters. We find
$P^{\prime}=0.573P$, and $\nu = 0.529$, shown in
Fig.~\ref{fig:f-actin_fd virus}(b), where we also compare our results
to the ideal Kratky-Porod worm-like chain, and to the rigid rod length
$L$. Clearly, for shorter micelles, $L_{e,I}$ approaches $L$, whilst
for longer ones, $L_{e,I}$ approaches the ideal Kratky-Porod $L_{KP}$.

In the nematic fd virus phase, we can consider the polymer using the
Odijk confined worm-like chain model \cite{ref:Odijk}, where the host
nematic acts as a confining cylinder. The average end-to-end length is
defined as $L_{O}=L\langle {\bf\omega}\cdot{\bf n}\rangle$,
proportional to the average ${\bf\omega}\cdot{\bf n}$ along the chain,
where ${\bf n}$ is the nematic director. We find that our calculated
effective length at the I-N phase transition $L_{e,N}$ is
significantly below $L_{O}$ (Fig.~\ref{fig:f-actin_fd virus}(c)), from
which we infer that the host nematic is neither dense nor ordered
enough to fully confine the polymer. By altering the stiffness of the
host fd virus such that it is denser (flexible fd virus) or more
ordered (rigid) at the transition (see the appendix), we may confine
better the polymer causing it to stretch out more over the I-N
transition. Fig.~\ref{fig:f-actin_fd virus}(d) shows $\Delta
L_{e}=L_{e,N}-L_{e,I}$ for worm-like micelles of $L=10\mathrm{\mu m}$,
using a range of $P$ values for the fd virus. Here we see that $\Delta
L_{e}$ is the largest when using the most flexible fd virus and it
decreases as the fd virus becomes more rigid, until it reaches a
minimum, before increasing again for the most rigid fd virus
studied. This presents the opportunity to tune the stretching of the
polymers by varying the stiffness of the host fd virus. We also
compare $\Delta L_{e}$ to $\Delta L^{\prime}_{e} = L_{O}-L_{KP}$
(Fig.~\ref{fig:f-actin_fd virus}(d)), finding that this overestimates
$\Delta L_{e}$, but does qualitatively match our non-monotonic
results.

In conclusion, we have developed a model for binary mixtures of
semi-flexible rods, and applied it to binary fd virus mixtures. We
find I-N, N-N (with upper, lower and no critical points) and I-N-N
coexistence regions present in such systems, consistent with
experimental results \cite{ref:Purdy}. Additionally, we have shown
that the N-N upper critical point only exists in a very small diameter
ratio regime, which may explain why it has never been seen
experimentally. For systems with diameter ratios larger than those
studied experimentally so far, we also find I-I and I-I-N coexistence
regions. We also see that altering the stiffness of the rods can have
the same effect on phase behavior as altering the diameter ratio,
giving an extra parameter to tuning the phase behavior of rod-like
particles. The addition of flexibility in our model gives results that
are quantitatively closer to experimental observations than those
obtained using rigid rods \cite{ref:Purdy}. The key reason appears to
lie in the effective shape of the rods, which changes throughout the
phase diagrams, indicating that any static-shaped rigid rod model will
miss some of the essential physics. We have also studied the
stretching of semi-flexible polymers in an fd virus solvent. We find
that sufficiently short polymers stretch out considerably over the I-N
transition of the host solvent, while for the longer ones, some
stretching is observed, but the effect is much less
pronounced. Changing the stiffness of the host fd virus, such that the
density or nematic order at the I-N transition is increased, can
greatly increase the stretching effect. We hope that our findings will
stimulate further experimental explorations, and believe that
extensions of the theory to e.g. inhomogeneous states and
non-equilibrium phenomena are within reach of the present model.

Financial support of a FOM and a NWO-VICI grant is acknowledged.

\bibliography{main}
\clearpage

\onecolumngrid
\section{Appendix}
\subsection{Theory}\label{sec:theory}
\subsubsection{Free-energy minimization and phase diagrams}\label{sec:PD}
In the Onsager theory \cite{ref:Onsager}, the isotropic-nematic phase
transition of a one-component system of rigid-rods is driven by
competition between two entropies. One is similar to entropy of
mixing, arising from the mixing of particles of different
orientations. The other arises from the excluded volume interactions
of the particles. These are expressed via the Helmholtz free energy,
$F$, and for monodisperse rigid particles the solutions to this are
well known (see e.g. \cite{ref:Onsager1,ref:Onsager2}). Extending the
theory to polydisperse systems requires the addition of an extra
mixing term to the Helmholtz free energy
\cite{ref:vRoij,ref:Experi1}. Flexibility may be incorporated in
numerous ways (see
e.g. \cite{ref:wessels1,*ref:wessels2,ref:KS1,*ref:KS2,ref:Chen,ref:Dijkstra_Frenkel,ref:Odijk}),
and in this supplementary material, we build on the work of Wessels
and Mulder \cite{ref:wessels1,*ref:wessels2} to produce a model
describing binary systems of semi-flexible rods.

We consider a suspension of $N_i$ semi-flexible rods of species
$i=1,2$ with contour lengths $L_i$, in a volume $V$ at temperature
$T$. Following Wessels and Mulder \cite{ref:wessels1,*ref:wessels2} we
model a rod of species $i$ as a chain of $M_i$ rod-like segments of
length $l_i=L_i/M_i$ and diameter $D_i\ll l_i$. Denoting the
orientation of the $m$-th segment by a unit-vector ${\bf\omega}_m$
(with $1\leq m\leq M_i$), we write the bending energy of a chain of
species $i$ with orientation
${\bf\Omega}=\{{\bf\omega}_1,\dots,{\bf\omega}_{M_i}\}$ as

\begin{equation}
U_{i}({\bf\Omega})=\sum_{m=1}^{M_{i}-1}u_{i}({\bf\omega}_{m},{\bf\omega}_{m+1})=
-\frac{P_i}{l_i}\sum_{m=1}^{M_{i}-1}{\bf\omega}_{m}\cdot{\bf\omega}_{m+1},
\label{eq:U}
\end{equation}
where the stiffness is described in terms of the persistence length
$P_i$ \cite{ref:wessels1,*ref:wessels2}. Here and below we use thermal
energy units by setting $k_BT=1$. The state of the suspension is
characterized by the orientation distributions functions (ODFs)
$f_i({\bf\Omega})$, which satisfy the normalization condition $\int
d{\bf\Omega} f_i({\bf\Omega})=1$ where
$d{\bf\Omega}=\prod_{m=1}^{M_i}d{\bf\omega}_m$. Denoting the total
number of rods by $N=N_1+N_2$, the density by $\rho=N/V$, and the mole
fraction of species $i$ by $x_i=N_i/N$, we can write the variational
free-energy functional $F[f_1,f_2]$ of this system within an
Onsager-like second virial approximation as

\begin{eqnarray}
\frac{F}{N} &= &\ln(B\rho)-1+ x_{1}\ln x_{1}+ x_{2}\ln
x_{2} \label{eq:helmholtz_bi1}
\\ &+&\displaystyle\sum_{i=1}^{2}x_{i}\displaystyle \int
f_{i}({\bf\Omega})\Big(\ln(4\pi
f_{i}({\bf\Omega}))+U_{i}({\bf\Omega})\Big)d{\bf\Omega} \nonumber
\\ &+&\frac{\rho}{2}\displaystyle\sum_{i,j=1}^{2}x_{i}x_{j}\displaystyle
\int f_{i}({\bf\Omega})f_{j}({\bf\Omega}^{\prime}) K_{ij}({\bf
  \Omega},{\bf\Omega}^{\prime})d{\bf\Omega}d{\bf
  \Omega}^{\prime}.\nonumber
\end{eqnarray}
The first line of Eq.~(\ref{eq:helmholtz_bi1}) represents the
translational and the mixing ideal-gas contributions (with
$B=\frac{\pi}{4}D_{1}L_{1}^{2}$, a constant), the second line, the
orientation entropy and bending energy, and the third line the
excluded volume interactions given by

\begin{eqnarray}
K_{ij}({\bf\Omega},{\bf\Omega}^{\prime}) &=&
\sum_{m=1}^{M_{i}}\sum_{m^{\prime}=1}^{M_{j}}k_{ij}({\bf
  \omega}_{m},{\bf\omega}_{m^{\prime}})\nonumber
\\ &=&l_{i}l_{j}(D_{i}+D_{j})\sum_{m=1}^{M_{i}}\sum_{m^{\prime}=1}^{M_{j}}
|\sin\gamma({\bf\omega}_{m},{\bf\omega}_{m^{\prime}})|, \label{eq:B_mix}
\end{eqnarray}
with $\gamma({\bf\omega}_{m},{\bf\omega}_{m^{\prime}})=
\arccos({\bf\omega}_{m}\cdot{\bf\omega}_{m^{\prime}})$ the angle
between chain segments \cite{ref:wessels1,*ref:wessels2} $m$ and
$m^{\prime}$. The free-energy functional of
Eq.~(\ref{eq:helmholtz_bi1}) is a two-component generalization of the
one-component segmented-chain functional of
Ref.\cite{ref:wessels1,*ref:wessels2}, and for $M_i=1$ and $U_i\equiv
0$ it reduces to the Onsager functional for binary mixtures of rigid
rods \cite{ref:Vroege,ref:vRoij}.

At a given thermodynamic state point, the equilibrium ODFs minimize
$F$ and therefore satisfy the Euler-Lagrange equations $\delta
(F-\mu_i N_i)/\delta f_i({\bf\Omega})=0$ for $i=1,2$, with $\mu_i$ the
chemical potential-like Lagrange multiplier that ensures a proper
normalization. This gives rise to

\begin{eqnarray}
f_i({\bf\Omega})&=&\frac{\exp(-U_i({\bf\Omega})-V_i({\bf\Omega}))}{Q_i}; \label{sca}\\ 
V_i({\bf\Omega})&=&\rho\sum_{j=1}^{2} x_j\int K_{ij}({\bf\Omega},{\bf\Omega}^{\prime})
f_j({\bf\Omega}^{\prime})d{\bf\Omega}^{\prime}\label{scb},
\end{eqnarray}
where $V_i({\bf\Omega})$ can be seen as a self-consistent field acting
on all segments of a chain, and $Q_i$ is a partition function-like
normalization factor. Explicitly solving Eqs.~(\ref{sca}) and
(\ref{scb}) for state points of interest would be prohibitively
expensive computationally because of the high-dimensional angular
${\bf\Omega}$-grids that would be required in the case when $M_i\gg
1$. Instead, we formally evaluate the functional $F$ of
Eq.~(\ref{eq:helmholtz_bi1}) in its minimum by inserting the solutions
$f_i$ of Eqs.~(\ref{sca}) and (\ref{scb}) to find the equilibrium free
energy

\begin{equation}
\label{eq1:helmholtz_bin_eq}
\frac{F_{\mbox{\small eq}}}{N} = \ln(B\rho) -1 +
x_{1}\ln\frac{x_{1}}{Q_{1}} + x_{2}\ln\frac{x_{2}}{Q_{2}} -
\frac{1}{2}\rho\sum_{i,j}^{2}x_{i}x_{j}
\sum_{m=1}^{M_{i}}\sum_{m^{\prime}=1}^{M_{j}} \int
k_{ij}({\bf\omega},{\bf\omega}^{\prime})f_{i,m}({\bf\omega})f_{j,m^{\prime}}
({\bf\omega}^{\prime})d{\bf\omega}d{\bf\omega}^{\prime},
\end{equation}
where $f_{i,m}({\bf\omega})$ is the ODF of the $m$-th segment
($m=1,\dots,M_i$) of a chain of species $i=1,2$ defined by

\begin{equation}
f_{i,m}({\bf\omega}_{m}) =\int f_i({\bf\Omega})d{\bf\omega}_1\dots
d{\bf\omega}_{m-1}d{\bf\omega}_{m+1}\dots d{\bf\omega}_{M_i}.
\label{eq:f_mth}
\end{equation}
Eq.~(\ref{eq1:helmholtz_bin_eq}) implies that the thermodynamics does
{\em not} require the full solutions $f_i({\bf\Omega})$ but in fact
only the $M_i$ single-segment distributions $f_{i,m}({\bf\omega})$ and
the normalization factors $Q_i$, for which an efficient iterative
recursion scheme, that exploits the connectivity of the chain, can be
set up as follows.

Eqs.~(\ref{eq:B_mix}) and (\ref{scb}) allow us to write
$V_i({\bf\Omega})=\sum_{m=1}^{M_i}v_i({\bf\omega}_m)$ with the {\em
  same} selfconsistent field

\begin{equation}
\label{eq:vi}
v_{i}({\bf\omega}_{m}) =\rho \sum_{j=1}^{2}\sum_{m^{\prime}=1}^{M_{j}} x_{j}
\displaystyle
\int k_{ij}({\bf\omega}_{m},{\bf\omega}_{m^{\prime}})f_{j,m^{\prime}}
({\bf\omega}_{m^{\prime}})d{\bf\omega}_{m^{\prime}},
\end{equation}
for all segments of chains of the same species. As a consequence,
Eq.~(\ref{sca}) combined with Eq.~(\ref{eq:f_mth}) can be written as

\begin{equation}
f_{i,m}({\bf\omega}) =
\frac{1}{Q_{i}}q_{i,m}({\bf\omega})\exp[-v_{i}({\bf\omega})]q_{i,M-m+1}({\bf\omega}),
\label{eq:f_min_bin}
\end{equation}
with the partial-chain partition function

\begin{equation}
q_{i,m}({\bf\omega}_m) = \int \prod_{n=1}^{m-1}\exp[-v_i({\bf\omega}_n)-
u_i({\bf\omega}_n,{\bf\omega}_{n+1})]d{\bf\omega}_{n}
\label{qim}.
\end{equation}
In the formulation of Eq.~(\ref{eq:f_min_bin}) the $m$-th segment ODF
is seen as the statistical weight $\exp(-v_i({\bf\omega}))$ of that
segment in the (selfconsistent) field $v_i({\bf\omega})$, combined
with the weights $q_{i,m}$ and $q_{i,M-m+1}$ of the two sub-chains
from segment $m\pm 1$ to the two chain ends,
respectively. Interestingly, the connectivity of the chain allows us
to rewrite Eq.~(\ref{qim}) as the recursion relation

\begin{equation}
q_{i,m}({\bf \omega})=\int q_{i,m-1}({\bf\omega}^{\prime})
\exp[-v_{i}({\bf\omega^{\prime}})-
u_{i}({\bf\omega^{\prime}},{\bf\omega})]d{\bf\omega^{\prime}}
\label{eq:q_bin}
\end{equation}
such that a loop can be set up that (i) starts with a guess for
$v_i({\bf\omega})$ for $i=1,2$, (ii) solves for $q_{i,m}({\bf\omega})$
for all $i=1,2$ and $m=1,\dots M_i$ using Eq.~(\ref{eq:q_bin})
together with $q_{i,1}({\bf\omega})\equiv 1$, (iii) computes
$f_{i,m}({\bf\omega})$ and the normalization factor $Q_i$ from
Eq.~(\ref{eq:f_min_bin}), (iv) recalculates $v_i({\bf\omega})$ using
Eq.~(\ref{eq:vi}) and repeats (ii)-(iv) until convergence is
found. Note that this scheme {\em only} requires an angular grid for
${\bf\omega}$, which in the light of the azimuthal and up-down
symmetry of the isotropic and nematic phases of interest here, reduces
to a single grid for the polar angle $\theta\in(0,\pi/2)$. The {\em
  only} difference with Onsager-type theories for rigid-rod mixtures
is the additional calculation and storage of $q_{i,m}({\bf\omega})$
for $1\leq m\leq M_i$ here. With the ODFs known, $F_{\mbox{\small
    eq}}$ may be calculated from Eq.~(\ref{eq1:helmholtz_bin_eq}). The
osmotic pressure is then calculated from $\Pi=\rho^{2} \frac{\partial
  F_{eq}/N}{\partial\rho}$. For a binary system, phase behavior is
most easily analyzed using the Gibbs energy per particle,
$\text{\emph{\~{g}}}(x,\Pi)=\frac{F_{eq}}{N}+\frac{\Pi}{\rho}$. By
fixing $\Pi$, $\text{\emph{\~{g}}}$ may be calculated as a function of
$x_{2}$ (with $x_{1}=1-x_{2}$), and performing a common tangent
construction allows for the prediction of coexisting phases
\cite{ref:vRoij,ref:Experi1,ref:vRoij2}. For the special case of a
monodisperse system, we simply set $x_{2} = 0$.

We may also calculate the nematic order parameters, $S_{i,m}$, which
define the local order of the $m$-th segment

\begin{equation}
S_{i,m} = \displaystyle \int d{\bf\omega}
f_{i,m}({\bf\omega})P_{2}({\bf\omega}\cdot{\bf n}),
\label{eq:S_nem_m}
\end{equation}
where $P_{2}({\bf\omega}\cdot{\bf n})$ is the second Legendre
polynomial, and ${\bf n}$ is the nematic director. We define the
nematic order of a rod as the average nematic order along the chain

\begin{equation}
S_{i} = \frac{1}{M_{i}} \sum_{m=1}^{M_{i}} S_{i,m}.
\label{eq:S_nem}
\end{equation}

\subsubsection{Effective length}\label{sec:PD}
The calculation of the effective length goes as follows. We define the
mean square effective length $L^{2}_{e,i}$ as

\begin{eqnarray}
  L^{2}_{e,i} &=& l^{2}_{i}\displaystyle
  \sum_{m=1}^{M_{i}}\sum_{m^{\prime}=1}^{M_{i}}\langle{\bf\omega}_{m}\cdot{\bf
    \omega}_{m^{\prime}}\rangle\nonumber \\ &=& l^{2}_{i}\displaystyle
  \sum_{m=1}^{M_{i}}\sum_{m^{\prime}=1}^{M_{i}}\int
      ({\bf\omega}\cdot{\bf\omega}^{\prime})f_{i,m,m^{\prime}}({\bf\omega},{\bf\omega}^{\prime})
      d{\bf\omega}d{\bf\omega}^{\prime},
\label{eq:eff_length}
\end{eqnarray}
where we are summing the squares of the average length projections of
all chain segments $m^{\prime}$ along the director of all segments
$m$. Here, $f_{i,m,m^{\prime}}({\bf\omega},{\bf\omega}^{\prime})$ is
the pair orientational distribution function (PDF) defined by

\begin{equation}
f_{i,m,m^{\prime}}({\bf\omega},{\bf\omega}^{\prime})=\displaystyle\int
f_{i}({\bf\Omega})\delta({\bf\omega}_{m}-{\bf\omega})\delta({\bf\omega}_{m^{\prime}}-{\bf\omega}^{\prime})d{\bf\Omega},
\label{eq:f_nth}
\end{equation}
where we are integrating out all other degrees of freedom from
$f_{i}({\bf\Omega})$ except those of segments $m$ and
$m^{\prime}$. Note that
$f_{i,m,m^{\prime}}({\bf\omega},{\bf\omega}^{\prime})$ is the
probability that a chain of species $i$ is in a configuration with the
$m$-th and $m^{\prime}$-th segment having orientations ${\bf\omega}$
and ${\bf\omega}^{\prime}$, simultaneously. Inserting Eq.~(\ref{sca})
into Eq.~(\ref{eq:f_nth}), and using Eqs.~(\ref{eq:vi}) and
(\ref{qim}), we find that

\begin{equation}
f_{i,m,m^{\prime}}({\bf\omega},{\bf\omega}^{\prime}) =
\frac{1}{Q_{i}}q_{i,m}({\bf\omega})\exp[-v_{i}({\bf\omega})]Q_{i,m,m^{\prime}}({\bf\omega},{\bf\omega}^{\prime})\exp[-v_{i}({\bf\omega}^{\prime})]q_{i,M-m^{\prime}+1}({\bf\omega}^{\prime}),
\label{eq:pdf}
\end{equation}
with the same notation as before. Here
$Q_{i,m,m^{\prime}}({\bf\omega},{\bf\omega}^{\prime})$ is the partial
chain partition function that takes into account the effect of the
chain segments that link segment $m$ and segment $m^{\prime}$. For
neighboring segments $m^{\prime}=m+1$ we have
$Q_{i,m,m^{\prime}}({\bf\omega},{\bf\omega}^{\prime})=\exp[-u_{i}({\bf\omega},{\bf\omega}^{\prime})]$,
and for $m^{\prime}=m+2 \dots M$, it follows the recursion relation

\begin{equation}
Q_{i,m,m^{\prime}}({\bf\omega},{\bf\omega}^{\prime})=
\displaystyle\int
d{\bf\omega}^{\prime\prime}Q_{i,m,m^{\prime}-1}({\bf\omega},{\bf\omega}^{\prime\prime})\exp[-v_{i}({\bf\omega}^{\prime\prime})]\exp[-u_{i}({\bf\omega}^{\prime\prime},{\bf\omega}^{\prime})].
\label{eq:Qm_recursion}
\end{equation}
By construction, each pair orientation distribution function also
obeys the normalization condition $\int
f_{i,m,m^{\prime}}({\bf\omega},{\bf\omega}^{\prime})
d{\bf\omega}d{\bf\omega}^{\prime}=1$. As we already know the ODFs, and
hence $q_{i,m}({\bf\omega})$ and $v_{i}({\bf\omega})$, from our phase
diagram calculations, the calculation of the PDFs and $L_{e,i}$ is
relatively straight forward. We use $L_{e,i}$ to calculate the
diameter $D_{e,i}$ required for rigid rods to have the same excluded
volume as our flexible rods, at the same state point, obtaining the
effective shape of the rods.

\subsection{fd virus parameters and values}\label{sec:limit}
\begin{figure}[!h]
    \includegraphics*[width=0.39\columnwidth,height=0.6\columnwidth,angle=270]{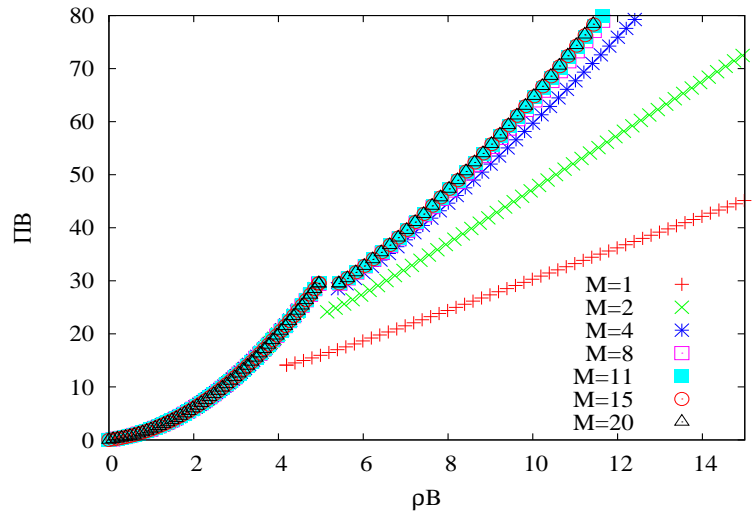}
    \caption{Convergence of equation of state in $\rho B - \Pi B$
      representation, for increasing number of chain segments $M$ for
      fd virus parameters $L=0.88\mathrm{\mu m}$, $D=6.6\mathrm{nm}$
      and $P=2.2\mathrm{\mu m}$. We find that $M=15$ is within the
      continuum limit for the fd virus.}
    \label{fig:EOS}
\end{figure}

In order to accurately describe semi-flexible rods, we must ensure
that we are in the continuum limit. That is, we must use a sufficient
number of chain segments to ensure that our results capture the
physics of a continuous chain. We do this by checking the convergence
of our results with increasing $M_{i}$ at fixed $P_{i}$, by adapting
$l_{i}=L_{i}/M_{i}$. We shall now examine the one-component system fd
virus system, and hence we drop the subscript $i$. In
Fig.~\ref{fig:EOS}, we show the isotropic (I) and nematic (N) equation
of state of a one-component fd virus system, for various $M$ values,
starting from the rigid-rod limit $M=1$. Clearly, the isotropic branch
is independent of $M$, while the nematic branch strongly depends on
$M$. The I-N coexistence, which is represented by the jump, shows a
phase transition that shifts to higher $\rho$ and $\Pi$ upon
increasing $M$, reaching a well-defined continuum limit in the
pressure regime of interest for $M\geq15$. Coexistence is found at
$\Pi B = 29.54$.

\begin{figure}[!h]
    \includegraphics*[width=0.4\columnwidth,height=0.6\columnwidth,angle=270]{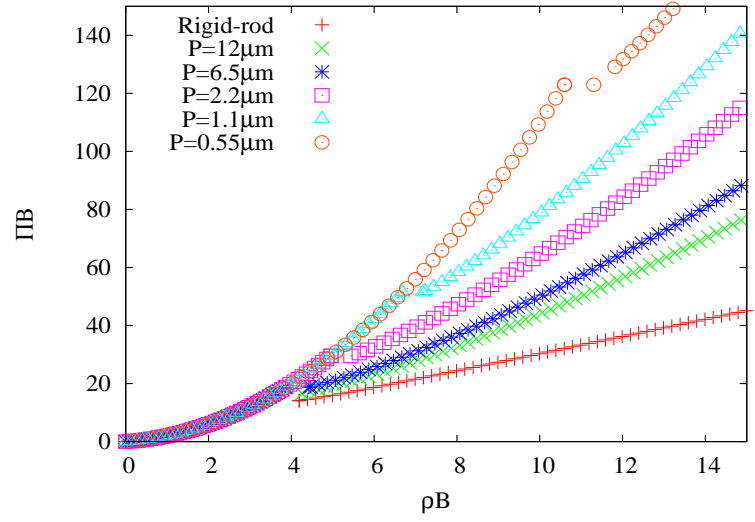}
    \caption{Equation of state in $\rho B - \Pi B$ representation, for
      fd virus dimensions $L=0.88\mathrm{\mu m}$, $D=6.6\mathrm{nm}$
      and varying persistence length $P$.}
    \label{fig:EOS2}  
\end{figure}

Varying the stiffness of semi-flexible rods can have a large effect on
the properties of the nematic phase at
coexistence. Fig.~\ref{fig:EOS2} shows the equation of state for
bio-engineered fd virus particles of various persistence lengths. We
see that coexistence is found at much lower densities (and osmotic
pressures) for stiff rods than for flexible ones, in agreement with
Refs.~\cite{ref:wessels1,*ref:wessels2,ref:KS1,*ref:KS2,ref:Chen,ref:Dijkstra_Frenkel,ref:Odijk}.

\begin{figure}[!h]
    \includegraphics*[width=0.4\columnwidth,height=0.6\columnwidth,angle=270]{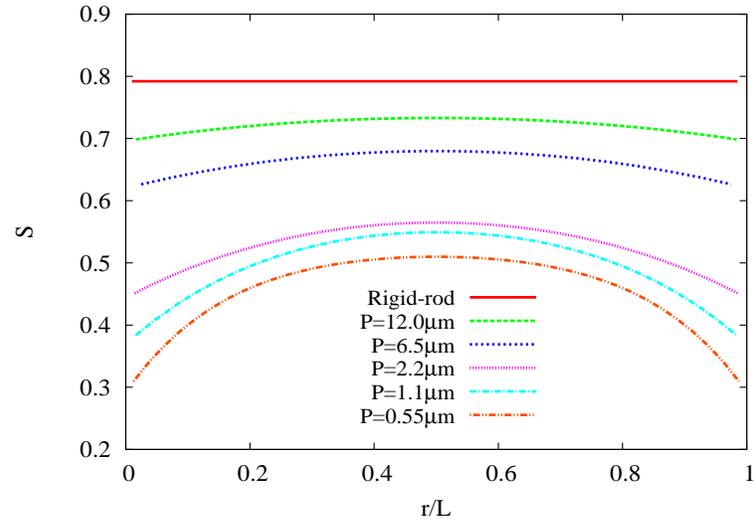}
    \caption{Nematic order parameter $S$ of fd virus particles of
      various persistence lengths, at a distance $r \in [0,L] $ along
      the rod. The densities of the nematic phase at coexistence are:
      $\rho B=4.19$, $\rho B=4.27$, $\rho B=4.42$, $\rho B=5.45$,
      $\rho B=7.17$ and $\rho B=11.30$ for $P\rightarrow \infty$
      (rigid-rod), $P=12\mathrm{\mu m}$, $P=6.5\mathrm{\mu m}$,
      $P=2.2\mathrm{\mu m}$, $P=1.1\mathrm{\mu m}$ and
      $P=0.55\mathrm{\mu m}$, respectively.}
    \label{fig:S2}
\end{figure}

We may also calculate the nematic order parameter of each chain
segment in a rod from Eq.~(\ref{eq:S_nem_m}). Fig.~\ref{fig:S2} shows
the nematic order parameter $S$ at a distance $r \in [0,L] $ along the
rod, at isotropic-nematic coexistence for fd virus particles of
various persistence lengths. We see that rigid rods are more ordered
at coexistence than flexible rods, despite coexistence being found at
a lower density. The chain segments in the middle of the rods are also
found to be more ordered than the end segments, in agreement with
earlier bifurcation findings in Ref.~\cite{ref:wessels1,*ref:wessels2}.

\end{document}